\documentclass[twocolumn,english]{IEEEtran}
\usepackage[T1]{fontenc}
\usepackage{fancyhdr}
\usepackage{graphicx}

\makeatletter

\providecommand{\tabularnewline}{\\}

\usepackage{cite}

\makeatother

\usepackage{babel}
\begin{document}

\title{A Compact CMOS Memristor Emulator Circuit and its Applications}

\author{{\normalsize{}Vishal Saxena, }\emph{\normalsize{}Member, IEEE}\thanks{V. Saxena is with the Department of Electrical and Computer Engineering,
University of Idaho, Moscow ID 83844 (e-mail: vsaxena@uidaho.edu)}}
\maketitle
\begin{abstract}
Conceptual memristors have recently gathered wider interest due to
their diverse application in non-von Neumann computing, machine learning,
neuromorphic computing, and chaotic circuits. We introduce a compact
CMOS circuit that emulates idealized memristor characteristics and
can bridge the gap between concepts to chip-scale realization by transcending
device challenges. The CMOS memristor circuit embodies a two-terminal
variable resistor whose resistance is controlled by the voltage applied
across its terminals. The memristor 'state' is held in a capacitor
that controls the resistor value. This work presents the design and
simulation of the memristor emulation circuit, and applies it to a
memcomputing application of maze solving using analog parallelism.
Furthermore, the memristor emulator circuit can be designed and fabricated
using standard commercial CMOS technologies and opens doors to interesting
applications in neuromorphic and machine learning circuits.
\end{abstract}

\begin{IEEEkeywords}
Memristors, memcomputing, neuromorphic, non-von Neumann computing,
ReRAM.
\end{IEEEkeywords}

\section{Introduction}

\IEEEPARstart{M}{}oore's Law has enabled the semiconductor industry
to sustain continual advancement in computing architectures by steadily
scaling transistor size, and adding more architectural complexity
while improving the energy-efficiency of the integrated circuit architectures.
This has sustained the development of several key computing and communication
technologies for several decades. However it appears that in order
to continue gaining from the transistor bounty of Moore's scaling
in the nanometer regime, we need to radically rethink existing computing
architectures. The new architectures are required to transcend the
device variability and interconnect scaling bottlenecks of the traditional
von-Neumann architecture, should exploit massive parallelism and locally
employ memory within the computing elements in a manner similar to
biological brains. One emerging technology that is promising for such
computing architectures is memristors, or resistive RAM (ReRAM). Recent
progress in memristive devices has spurred renewed interest in reconfigurable
and neuromorphic computing architectures \cite{strukov2008missing,rothenbuhler2013reconfigurable,vourkas2016memristor,kuzum2011nanoelectronic,jo2009high}. 

The memristive devices, integrated with conventional CMOS, are expected
to realize low-power neuromorphic circuits with increased reconfigurability
and smaller physical layout area \cite{indiveri2013integration,strukov2006reconfigurable}.
Since multi-valued memristor device technology is yet to mature and
its analog behavior in-situ with circuits is still being characterized
in the literature\cite{wu2015homogeneous,wu2016enabling,saxena2017towards},
we propose a low-risk and robust alternative for prototyping machine
learning and neuromorphic systems in standard CMOS. The rest of the
manuscript is organized as follows. Section II presents a brief background
on memristors along with the defining characteristics of memristive
circuits and devices. Section III presents the CMOS memristor (emulator)
circuit along with simulation results. Finally Section IV presents
an application using the proposed circuit, followed by conclusion.

\section{Memristor Characteristics}

Memristor was defined as a two-terminal circuit-theoretic concept
in a seminal paper by Chua \cite{chua1971memristor}, and later extended
to a wider class of memristive systems \cite{chua1976memristive}.
The generic equation to describe such memristive systems is given
by

\begin{equation}
y=g(x,u,t)\cdot u\label{eq:mem1}
\end{equation}

\begin{equation}
\dot{x}=f(x,u,t)\label{eq:mem2}
\end{equation}
where $x$ represents the state, $u$ and $y$ are the outputs of
the system respectively, $f$ is a continuous function n-valued function,
and $g$ is a scalar function \cite{radwan2015memristor,vourkas2016memristor}.
This relation can be interpreted as a voltage-controlled memristor,
where the current ($i$) in the memristor is related to the voltage
($v$) by 
\begin{equation}
i=G(x)\cdot v\label{eq:mem3}
\end{equation}
with its state update equation
\begin{equation}
\dot{x}=f(x,v,t)\label{eq:mem4}
\end{equation}

Here, $G(x)$ is called memductance (i.e. memory conductance), and
depends upon the charge that has flown across the device \cite{vourkas2016memristor}.
Decades later in year 2008, HP correlated the ideal memristor concept
with a metal-oxide resistive switching thin-flim device \cite{strukov2008missing}.
In the same paper, a linear ion drift model was introduced to describe
the behavior of the memristive device. Since then, several models
have been investigated to fit the equations to the experimental device
behavior \cite{joglekar2009elusive,yakopcic2011memristor,vourkas2016memristor}.
A unique set of `fingerprints' have been established based on the
underlying circuit theory, to distinguish memristors from other resistance-switching
devices. A memristor must exhibit a `pinched' hysteresis loop that
must pass through the origin in its $i-v$ switching curve, when the
applied sinusoid input is zero \cite{chua2011resistance}. Further,
the pinched hysteresis loops of ideal memristors must be odd symmetric,
otherwise they are modeled as generic memristive devices \cite{chua2011resistance}.
 Another fingerprint of a memristor is that, as the frequency of
the input sinusoid is increased, the area enclosed in the pinched
loops progressively shrinks and eventually collapses to a straight
line, or a single-valued function \cite{vourkas2016memristor}. The
fundamental promise of the memristors lies in the `analog' memory,
that endows it with the ability to store as well as manipulate information
in analog-domain, opposed to digital very large-scale integrated circuits
(VLSICs). Furthermore, these elements can be combined to realize analog
computing parallelism, enshrined under \emph{memcomputing} \cite{pershin2011solving,vourkas2016memristor,traversa2017polynomial}.
This analog computing ability is also harnessed in neuromorphic computing,
where memristors realize analog synapses that learn based on spike-timing
dependent plasticity (STDP), a local computing rule that is being
investigated as an unsupervised learning approach for deep learning
\cite{indiveri2013integration,Serrano-Gotarredona2013,wu2015cmos,wu2015homogeneous,neftci2016event,saxena2017towards}.

\section{CMOS Memristor Emulator Circuit}

In-spite of several promising features of idealized memristors, their
large-scale adoption has been impeded by the practical limitations
of fabricated memristive or resistive-RAM (ReRAM) devices. Device
characteristics such as the switching threshold voltages and resistances
are variable (and stochastic) for each device and across several devices,
and depend upon the initial \textquoteleft forming\textquoteright{}
step \cite{waser2009redox,yu2011stochastic,ielmini2015resistive}.
Further, it is challenging to realize stable weights for more than
1-bit resolution in filamentary devices due to the relaxation of the
filament and is currently being addressed in device research \cite{liu2013analog,saxena2017towards}.
HfOx and TaOx based devices have exhibited up to 9 states and their
performance within a circuit is being investigated \cite{beckmann2016nanoscale}.
A greater impediment from circuit design perspective is the lower
on-resistance observed in reported devices ($100\Omega-100k\Omega$),
which leads to power hungry driver circuits, and thus a fundamental
trade-off between parallelism and energy-efficiency. Moreover, memristive
devices fall well short of CMOS components in terms of their endurance.
Thus, it is desirable to realize CMOS memristor (emulator) circuits
for system-level design memcomputing exploration while the memristive
devices mature and are integrated into a commercially available CMOS
platform. 
\begin{figure}[h]
\begin{centering}
\includegraphics[width=0.9\columnwidth]{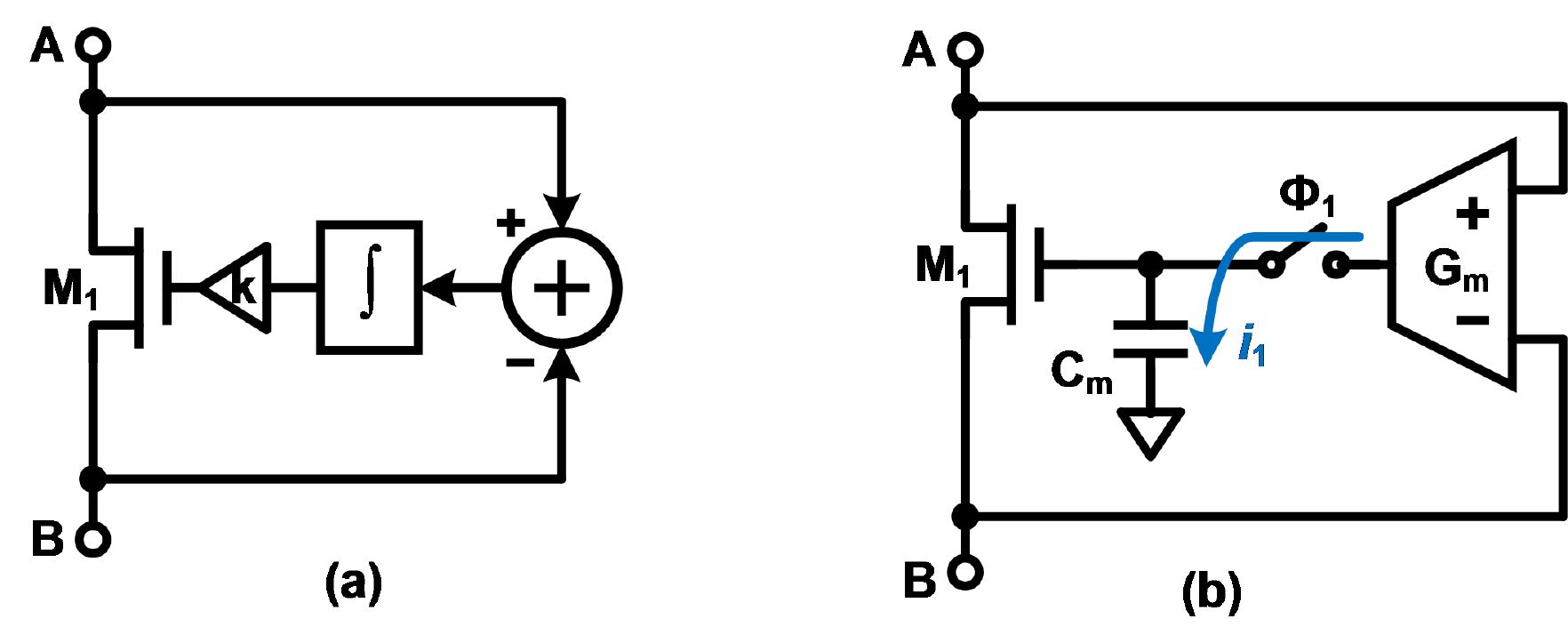}
\par\end{centering}
\caption{\label{fig:(a)-Conceptual-block}(a) Conceptual block diagram for
the CMOS memristor emulator, (b) A circuit implementation of the memristor
concept.}
\end{figure}

We disclosed the dynamic memristor/synapse circuit concept in the
patent application \cite{saxena2015memory}. In this work, we present
CMOS memristor circuit design details, analysis, and its application
in memcomputing. The fundamental concept is illustrated in Fig. \ref{fig:(a)-Conceptual-block}
(a), where an n-channel MOSFET (NMOS) $M_{1}$ implements a floating
variable resistance between terminals A and B. The variable resistance
is achieved by operating the transistor $M_{1}$ in linear (triode)
or near-linear region. The voltage difference across the terminals
A and B, $V_{AB}$, is sensed, integrated over time and then used
to control the gate of the transistor $M_{1}$. The integrator holds
the `state' of the variable resistor and updates it according to the
voltage difference $V_{AB}$. If a positive voltage is applied across
the resistor, the gate voltage is increased and thus resulting in
a decrease in the resistance (or an increase in conductance) across
A and B. Similarly, a negative potential across the resistor results
in an increase in the resistance (or decrease in conductance). This
circuit embodying a variable resistor with a dynamic memory, realizes
the functionality of the idealized memristor concept. Further, this
compact circuit can be modified to implement analog synapses that
exhibits bio-compatible learning rules, including the spike-timing
dependent plasticity (STDP) \cite{saxena2015memory}.
\begin{figure}[h]
\begin{centering}
\includegraphics[width=0.5\columnwidth]{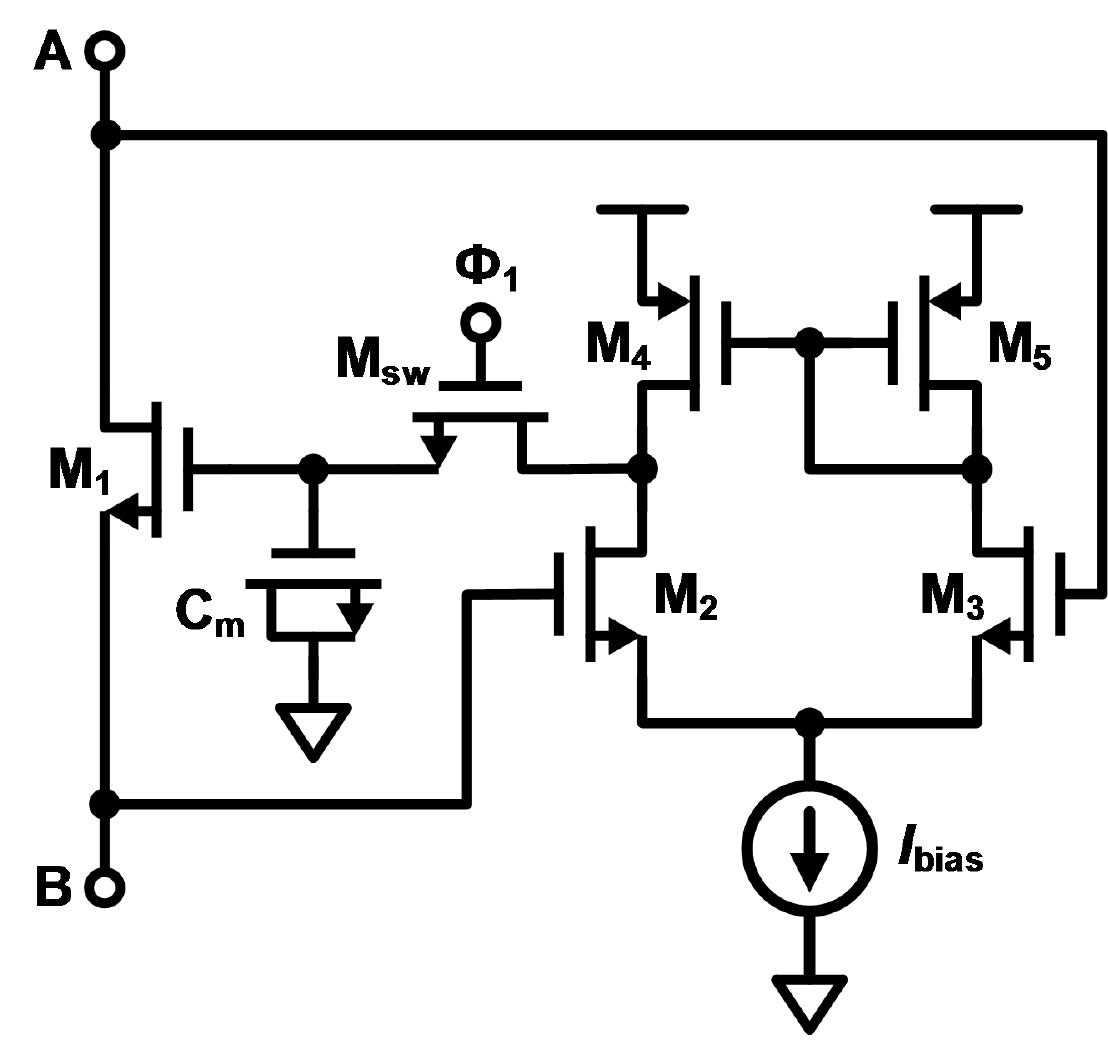}
\par\end{centering}
\caption{\label{fig:A-transistor-level-realization}A transistor-level realization
of the memristor (emulator) seen in Fig. \ref{fig:(a)-Conceptual-block}. }
\end{figure}

\begin{table}[h]
\caption{\label{tab:Design-parameters-for}Design parameters for memristor
circuit (Fig. \ref{fig:A-transistor-level-realization}) designed
in 130nm CMOS with $L_{min}=0.12\mu m$ and $V_{DD}=1.2V$. }
\centering{}%
\begin{tabular}{|c|c|c|}
\hline 
{\footnotesize{}Device} & {\footnotesize{}Type} & {\footnotesize{}Sizing (W/L) or value}\tabularnewline
\hline 
\hline 
{\footnotesize{}$M_{1}$} & {\footnotesize{}ZVT NMOS } & {\footnotesize{}$3\mu m/0.42\mu m$}\tabularnewline
\hline 
{\footnotesize{}$C_{m}$} & {\footnotesize{}NMOSCAP} & {\footnotesize{}$100fF$, $2.6\mu m\times2.6\mu m$}\tabularnewline
\hline 
{\footnotesize{}$M_{2}$} & {\footnotesize{}NMOS} & {\footnotesize{}$0.16\mu m/0.12\mu m$}\tabularnewline
\hline 
{\footnotesize{}$M_{3}$} & {\footnotesize{}NMOS} & {\footnotesize{}$0.16\mu m/0.12\mu m$}\tabularnewline
\hline 
{\footnotesize{}$M_{4}$} & {\footnotesize{}PMOS} & {\footnotesize{}$1.2\mu m/0.12\mu m$}\tabularnewline
\hline 
{\footnotesize{}$M_{5}$} & {\footnotesize{}PMOS} & {\footnotesize{}$1.2\mu m/0.12\mu m$}\tabularnewline
\hline 
{\footnotesize{}$M_{sw}$} & {\footnotesize{}NMOS} & {\footnotesize{}$1.2\mu m/0.12\mu m$}\tabularnewline
\hline 
{\footnotesize{}$I_{bias}$} &  & {\footnotesize{}$100nA$}\tabularnewline
\hline 
\end{tabular}
\end{table}

Fig. 1(b) shows one of several possible circuit implementations of
a voltage-controlled memristor, and its transistor-level implementation
is shown in Fig. \ref{fig:A-transistor-level-realization}. The circuit
was implemented in a 130-nm CMOS process with a supply voltage of
$V_{DD}=1.2V$, and the extracted layout was simulated in Cadence
Spectre. The corresponding device sizes are listed in Table \ref{tab:Design-parameters-for}.
Here, the transconductor $G_{m}$ senses the voltage across the two
terminals, produces a small-signal current which is integrated into
the charge stored on capacitor $C_{m}$, with an effective gain $k$.
The current integration phase can be controlled by an external strobe
or clock signal $\Phi_{1}$, whose utility will be apparent later
in this section. When $\Phi_{1}$ is $0V$, the capacitor $C_{m}$
is disconnected from the transconductor and holds the stored charge;
thus realizing a dynamic analog memory. The value of $C_{m}$ can
be in the range of $10\,fF$-$1\,pF$ in CMOS implementation, and
the ratio $\frac{I_{bias}}{C_{m}}$ determines the pinched hysteresis
characteristics (i.e. the width of the lobes) for a given input frequency,
$f_{in}$. Here, the voltage across the capacitor, $V_{G}$, controls
the gate of $M_{1}$ and referred to as the \textquoteleft state\textquoteright{}
of the synapse. Assuming that the transistor $M_{1}$ is in deep-triode,
the current flowing through the synapse is given by

\begin{equation}
I\approx KP_{n}\frac{W}{L}(V_{GS}-V_{THN})\cdot V_{AB}\label{eq:-1}
\end{equation}

where $V_{GS}$ and $V_{THN}$ are the gate-to-source and threshold
voltages, and $KP_{n}$ is the transconductance parameter, $W$ is
the width and $L$ is the length of the NMOS $M_{1}$. If the source
of $M_{1}$ is held at a common-mode voltage $V_{CM}$, the above
equation can be equivalently interpreted as the variable conductance
($G$)

\begin{equation}
G\approx KP_{n}\frac{W}{L}(V_{G}-V_{CM}-V_{THN})\label{eq:-3}
\end{equation}

which expresses a direct relation between the state ($x\equiv V_{G}$)
and the conductance ($G$). In order to force $M_{1}$ in triode for
large drain-source voltage swings, a zero-threshold voltage transistor
(ZVT) device ($V_{THN}=0$) available in standard CMOS platforms is
preferred. Also, in these scaled CMOS technologies, the MOSFET output
resistance ($r_{o}=g_{ds}^{-1}$) in moderate saturation is inherently
low due to short channel effects. This allows the proposed circuit
to exhibit desired memristive behavior even when $M_{1}$  is in moderate
saturation. In general, the dynamics of the memristor emulator circuit
can be described as 

\begin{equation}
I=G(x)\cdot V_{AB}\label{eq:mem3-1}
\end{equation}
\begin{equation}
\dot{x}=\frac{G_{m}(V_{AB})}{C_{m}}\cdot V_{AB}\equiv f(V_{AB},t)\label{eq:mem4-1}
\end{equation}

where Eqs. \ref{eq:mem3-1} and \ref{eq:mem4-1} are equivalent to
memristor equations seen in Eqs. \ref{eq:mem3} and \ref{eq:mem4}
respectively. 

\begin{figure}[tb]
\begin{centering}
\includegraphics[width=1\columnwidth]{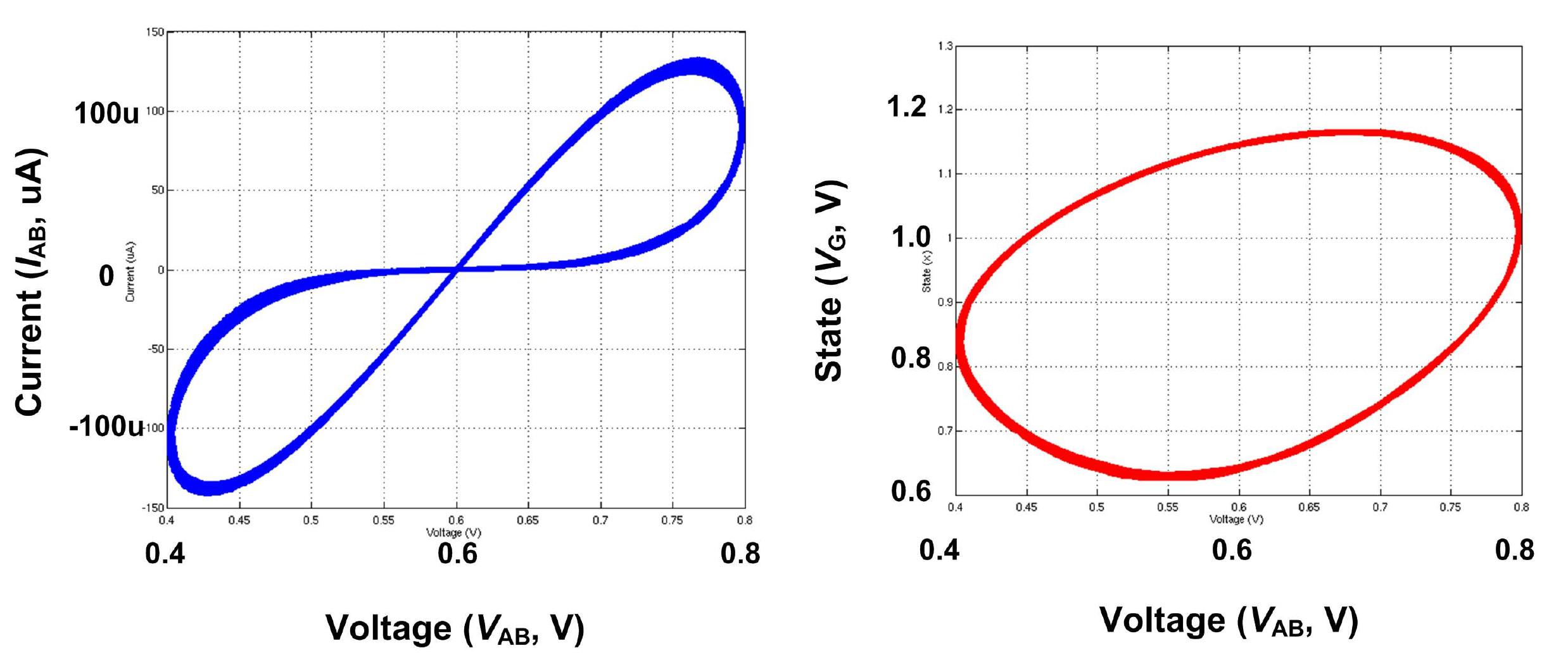}
\par\end{centering}
\caption{\label{fig:IV_curves_1}(a) Current-voltage hysteresis curve for the
a synapse circuit, (b) State trajectory for the synapse when swept
with a sine-wave input at $f_{in}=1MHz$. }
\end{figure}
\begin{figure}[h]
\begin{centering}
\includegraphics[width=1\columnwidth]{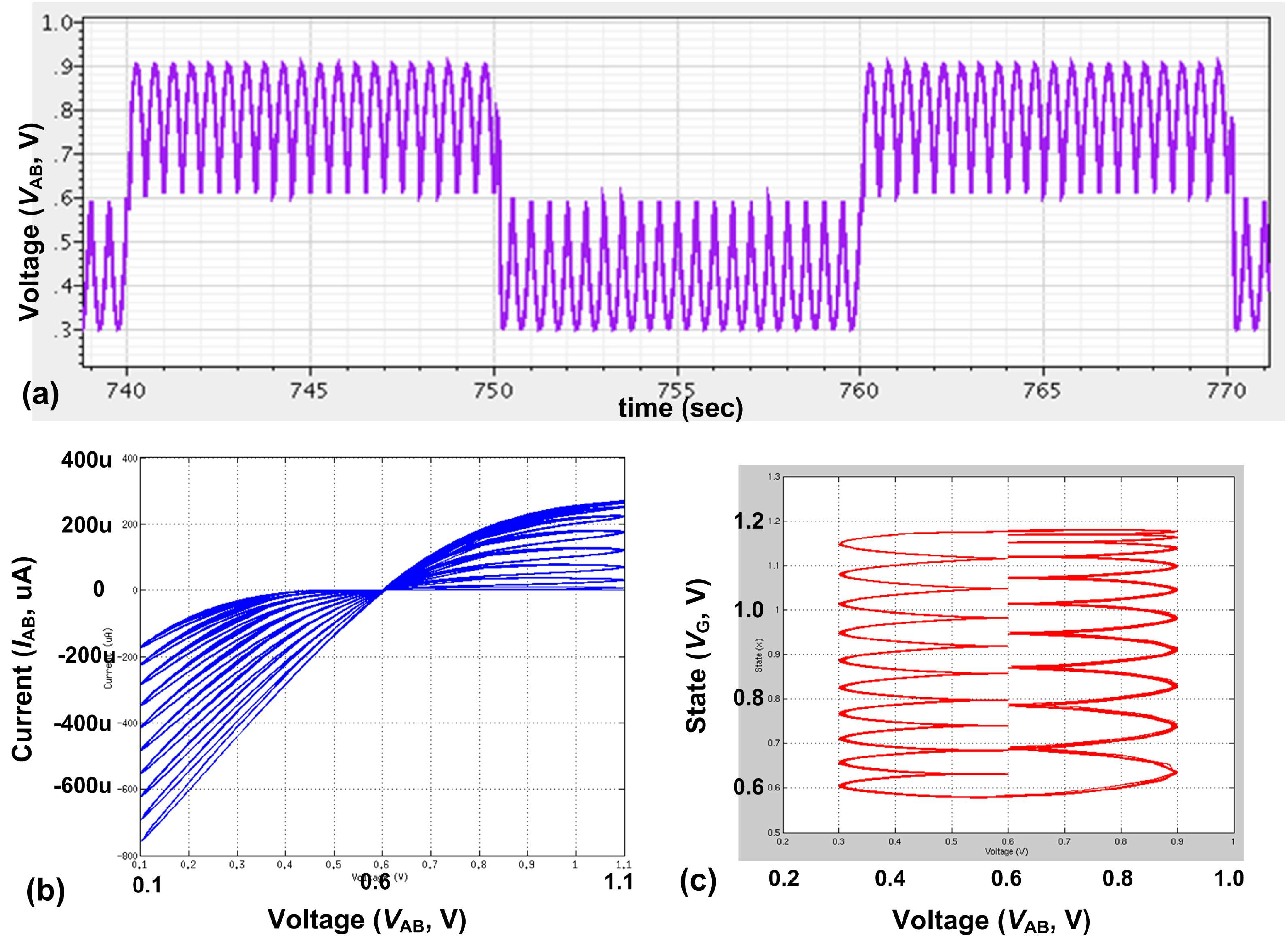}
\par\end{centering}
\caption{\label{fig:IV_curves_2}(a) Modulated half-sine wave input waveform
to generate successive voltage sweeps, (b) I-V curves with the modulated
sine-wave input at 1 GHz frequency, (c) state trajectory for the synapse.}
\end{figure}

 The simulated current-voltage hysteresis curve for the memristor
circuit is shown in Fig. \ref{fig:IV_curves_1}. The memristor was
swept with a sinusoidal input of frequency 1 MHz, amplitude 200mV
and a common-mode DC offset $V_{CM}$ equal to 600mV, while $\Phi_{1}$
was held at logic high (1.2 V). The circuit characteristics in Fig.
\ref{fig:IV_curves_1}(a) exhibit a pinched hysteresis curve typical
of an ideal memristor. The memory effect in the synapse is further
confirmed by the elliptical-like state trajectory seen in Fig. \ref{fig:IV_curves_1}(b).
The memristor circuit is further characterized by applying a modulated
half sine-wave (derived from a sine-wave of 1 GHz frequency) as shown
in Fig. \ref{fig:IV_curves_2}(a). This waveform can be understood
as successive voltage sweeps being applied to the synapse, to trace
closely spaced analog memory states. The circuit characteristics in
Fig. \ref{fig:IV_curves_2}(a) further exhibit near-ideal memristor
behavior with multiple pinched lobes. The synapse state trajectory
seen in Fig. \ref{fig:IV_curves_2}(b), which demonstrates that the
synapse can hold fine-grained analog memory states that can be precisely
controlled by an external stimulus. Moreover, the pinched lobes collapse
into a straight line as the input frequency is increased, as a signature
of memristors as discussed in Section II. An advantage of this circuit
is that the resistance range of the emulated memristor can be chosen
as desired depending upon the sizing of NMOS $M_{1}$; a flexibility
not conveniently available with memristive devices.

The memristor circuit in Fig. \ref{fig:A-transistor-level-realization}
inherently realizes a current-mode sample-and-hold to hold the analog
state of the synapses. The sampling switch ($M_{sw}$) prevents the
transconductor $G_{m}$'s output from leaking the state $(V_{G}$)
on capacitor $C_{m}$ when no inputs are applied (since $G_{m}$'s
output impedance is finite). This operation can also be interpreted
as an energy barrier which prevents the stored electrons on $C_{m}$
from leaking-out immediately. The electrons will however leak-out
eventually due to subthrehsold and junction leakages in the MOSFETs.
However, we can achieve storage times of up to several seconds by
using large $C_{m}$ and a low-leakage devices. Fig. \ref{fig:A-transistor-level-realization}
shows transient simulation results which exhibit the analog state
manipulation and short-term retention property of the dynamic memristor.
Here, a pseudo-random sequence is applied with ($\Delta T$ =5 ns,
$V_{pulse}=100mV$ amplitude, $V_{CM}=600mV$ DC offset) and 1 MHz
clock rate. We can observe that the state, $V_{G}$, is incrementally
updated corresponding to a positive or negative pulse input. Since
the conductance updates of the memristor are monotonic with respect
to the applied pulses (or voltage spikes), it can be used to dynamically
store analog weights in a machine learning or spiking neural network
circuit \cite{wu2015homogeneous}. The drain-source resistance of
$M_{1}$ is altered by updating the gate voltage, $V_{G}$, by applying
spikes across the terminals A-B. Assuming, rectangular spiking pulses,
the incremental update in the gate voltage can be expressed as 

\begin{equation}
\Delta V_{G}\approx\frac{G_{m}}{C_{m}}\cdot V_{spk}\Delta T\label{eq:}
\end{equation}

where $V_{spk}$ and $\Delta T$ are the spike height and width respectively.
\begin{figure}[h]
\begin{centering}
\includegraphics[width=0.8\columnwidth]{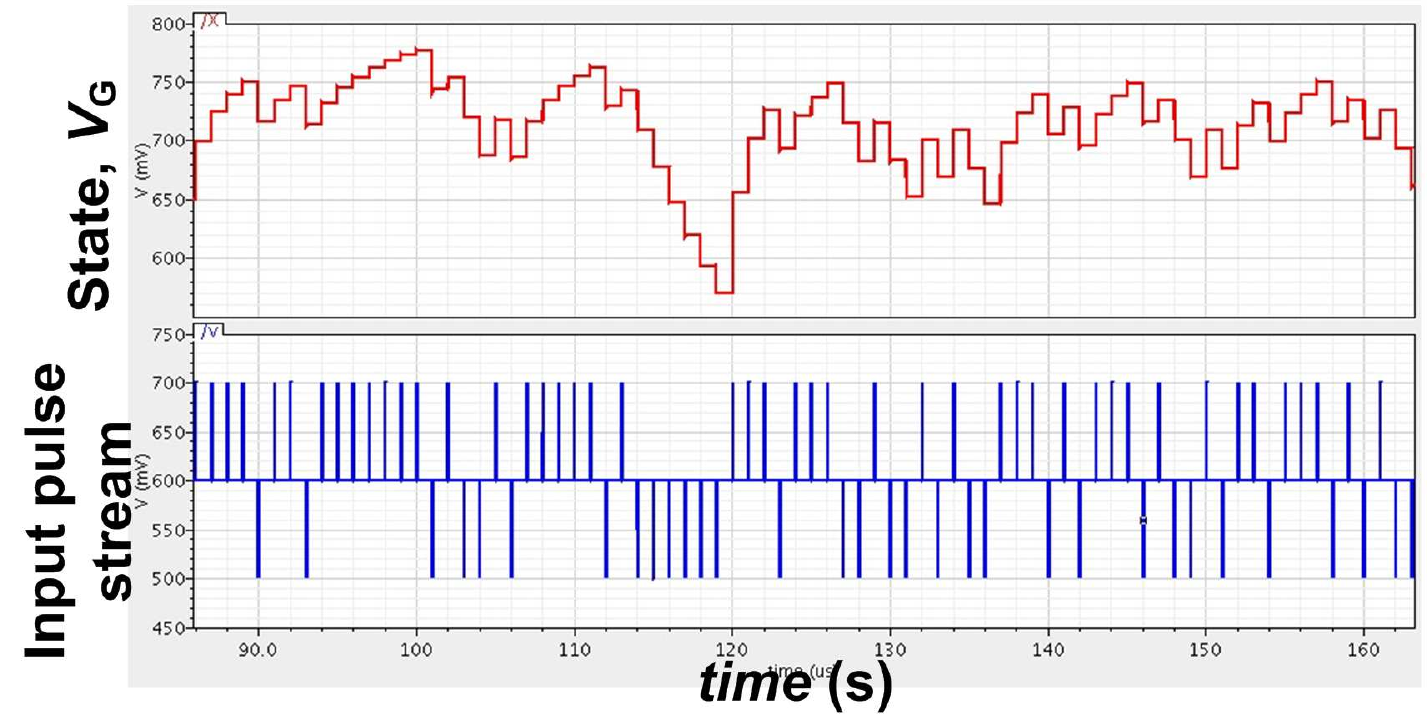}
\par\end{centering}
\caption{\label{fig:Pulsed-characterization-response}Pulsed characterization
response of the synapse circuit showing the monotonic incremental
control of the state, when positive and negative pulse are applied. }
\end{figure}

The presented memristor circuit was conceived while trying to understand
the relation between the pinched hysteresis behavior of memristor
and their ability to hold analog states. As shown in this work, the
switch $M_{sw}$ is not needed for generating the pinched hysteresis
sweep, but is essential for retaining the analog state by presenting
a barrier for the charge to leak away. This reinforces the understanding
that in order to realize analog memristive device, an electrochemical,
tunneling\cite{nandakumar2016250} or other form of barrier is needed
to retain multiple analog-like states. Also, in the presented dynamic
memristor, the state eventually leaks away. A bistable version of
the circuit can be realized by incorporating a weak latch structure
to quantize the dynamic state for long-term binary retention (see
\cite{saxena2015memory}). 

The transconductor-based memristor emulator circuit presented in this
work, even though simple in structure, offers tremendous advantages
over previously reported emulators \cite{fitch2013development}. The
circuit is compact, consumes very small bias current ($10-100nA$),
can be fabricated in standard CMOS technology, offers design flexibility
to choose memristor specifications, and doesn't require an analog-to-digital
converter (ADC)\cite{pershin2010practical}, or opamps \cite{muthuswamy2010simplest,valsa2011analogue,kim2012memristor,shin2013small,sanchez2014floating,yang2014memristor}
to implement its functionality.

\section{Memcomputing Application}

\begin{figure}[tb]
\centering{}\includegraphics[width=1\columnwidth]{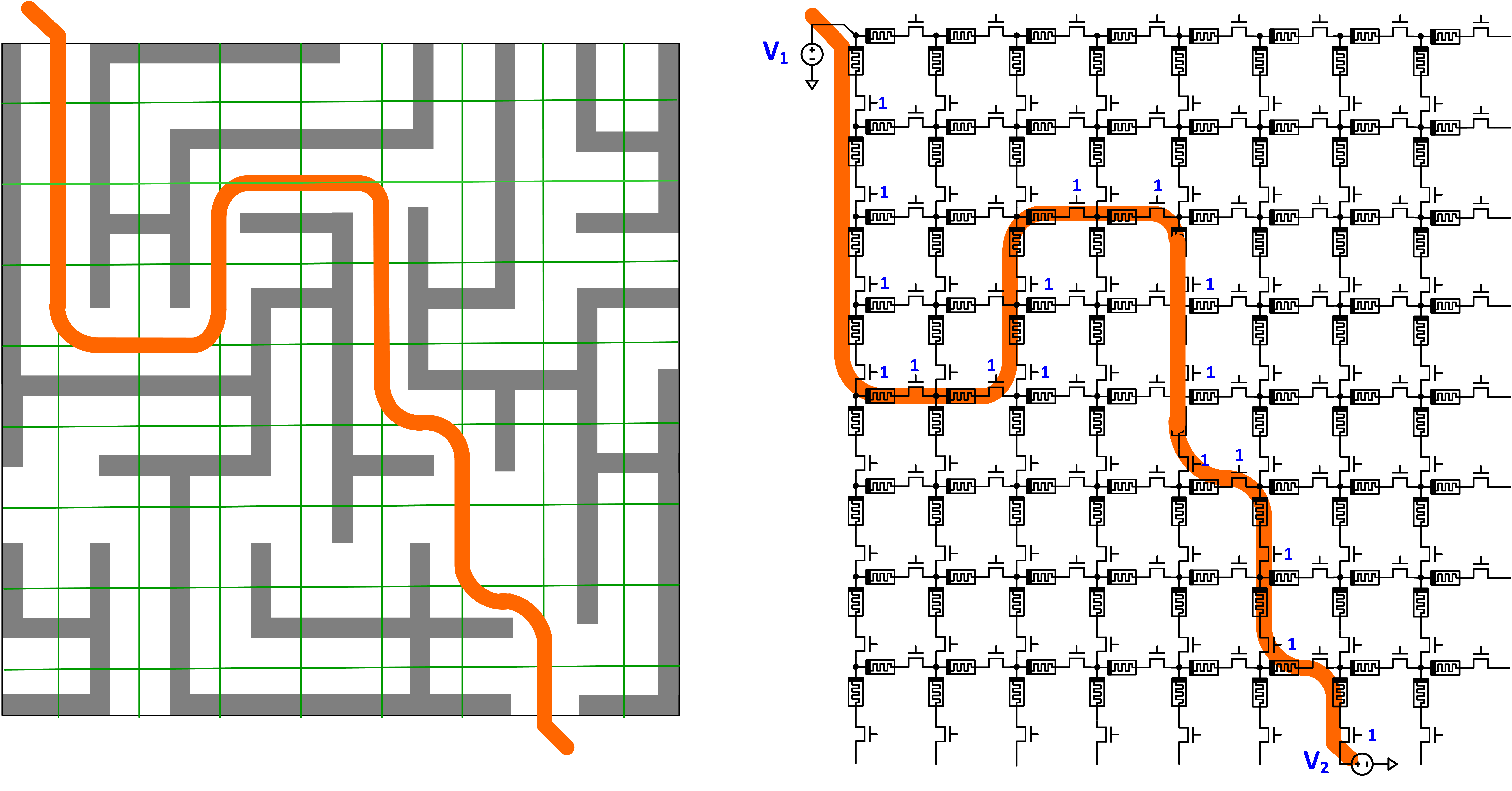}\caption{\label{fig:The--maze}The $8\times8$ maze used for simulation using
the CMOS memristor network. }
\end{figure}

\begin{figure}[tb]
\centering{}\includegraphics[width=1\columnwidth]{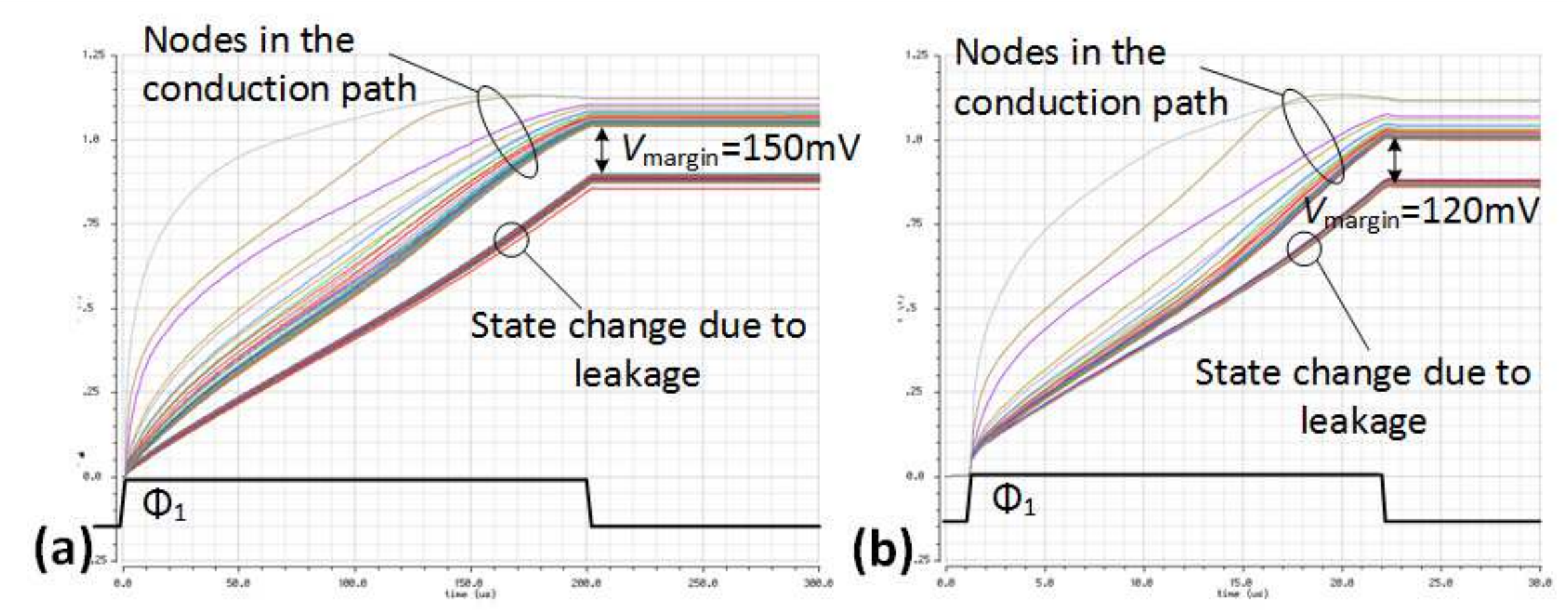}\caption{\label{fig:Simulation-results-for}Simulation results for the maze
with: (a) $C_{m}=1pF$ and $I_{bias}=1\mu A$, (b) (left) $C_{m}=100fF$
and $I_{bias}=100nA.$}
\end{figure}

The developed memristor emulator circuit can be leveraged for system-level
exploration of memcomputing and neuromorphic computing applications
\cite{pershin2011solving,traversa2017polynomial}, in parallel to
memristive device development. Here, we apply the circuit to realize
a 2D maze solver using a memristive network based on \cite{pershin2011solving}
and is shown in Fig. \ref{fig:The--maze}. Each tile of this network
consists of two CMOS memristors and two NMOS switches each. This $8\times8$
grid size network topology is generic and can be configured to a specific
maze pattern by selectively opening and closing the switches in Fig.
\ref{fig:The--maze} \cite{pershin2011solving,vourkas2016memristor}.
Initially, the states of all the memristor cells are set to zero (or
high-resistance state) in simulation by initializing the respective
gate voltages ($V_{G}$) to 0V. The array addressing scheme for switch
configuration and state reset is not shown here. DC voltages $V_{1}=800mV$
and $V_{2}=400mV$ are applied across the desired entrance and exit
nodes in the maze to solve for available paths, and the strobe $\Phi_{1}$
is set high to allow the network to settle. The simulation results
for this network are illustrated in Fig. \ref{fig:Simulation-results-for},
where the state voltages for each of the memristor cells are plotted
as a function of time, after the strobe $\Phi_{1}$ is applied. Here,
the memristors in the path connecting the entrance and the exit, i.e.
the solution of the maze, start conducting current as they decrease
their respective resistances. On the other hand, the memristors in
the non-conducting paths also start conducting due to leakage in the
memristor circuit and the `sneak paths', and their states increase
as well, but at a lower rate as seen in Fig. \ref{fig:Simulation-results-for}
(a\&b). Eventually all state nodes try to reach closer to $V_{DD}$.
The result of the network is read out by setting the strobe period
such that the states of the `On' memristors saturate, allowing for
a voltage margin ($V_{margin}$) for sensing as shown in Fig. \ref{fig:Simulation-results-for}.
The nodes that are in the maze solution path can be determined by
thresholding their states above the voltage level set by leakage currents
in `Off' memristors. Here, the static power consumption is around
$12.9\mu A$ and peak dynamic current is \textasciitilde{}$4.8\mu A$
from the supply. This application circuit can be further optimized
for maximizing the read margin, and the decision threshold and strobe
times can be digitally calibrated to compensate for process, voltage
and temperature (PVT) variations. 

\section{Conclusion\label{sec:Conclusion}}

A compact memristor emulator circuit in standard CMOS has been presented
along with simulation results. The circuit exhibits the signature
characteristics for ideal memristors and can transcend the challenges
associated with memristive device development in the near future.
Furthermore, the circuit has been applied to parallel maze solving,
and paves the path for applying the concept to other NP-hard problems
that can be solved by exploiting the parallel analog computation. 

\bibliographystyle{IEEEtran}
\bibliography{Memristors}

\end{document}